\documentclass[pra,twocolumn,showpacs,floatfix]{revtex4-1}
\usepackage{times,amsmath,amssymb,amstext,latexsym,float,graphicx,color}
\usepackage[hidelinks]{hyperref}
\hypersetup{colorlinks=true, citecolor=blue, urlcolor=blue, linkcolor=blue}
\newcommand{\kb}{k_{B}}
\newcommand{\lt}{L}
\newcommand{\xins}{\ell^{\mathrm{ins}}}

\begin{document}
\title{\bf Supremacy of the quantum many-body Szilard engine with attractive bosons }

\author{J.~Bengtsson$^{1,2}$, M.~Nilsson~Tengstrand$^{1,2}$, A. Wacker$^{1,2}$, P. Samuelsson$^{1,2}$, M.~Ueda$^{4,5}$, H. Linke$^{1,3}$  \& S.~M.~Reimann$^{1,2}$}

\affiliation{$^1$NanoLund, Lund University, P.O.Box 118, SE-22100 Lund, Sweden\\
$^2$ Mathematical Physics, Lund University, Box 118, 22100 Lund, Sweden\\
$^3$ Solid State Physics, Lund University, Box 118, 22100 Lund, Sweden\\
$^4$Department of Physics, University of Tokyo, 7-3-1 Hongo, Bunkyo-ku, Tokyo 11
3-0033, Japan\\
$^5$RIKEN Center for Emergent Matter Science (CEMS), Wako, Saitama 351-0198, Japan}

\date{\today}

\begin{abstract}
In a classic thought experiment, Szilard~\cite{Szilard1929} suggested a heat engine 
where a single particle, for example an atom or a molecule, is confined in a container coupled to a single heat bath. 
The container can be separated into two parts by a moveable wall acting as a piston.  
In a single cycle of the engine, work can be extracted from the information on 
which side of the piston the particle resides. The work output  
is consistent with Landauer's principle that the erasure of one bit of 
information costs the entropy $k_B \ln 2$~\cite{Landauer1961,Bennett1982,Leff2003}, 
 exemplifying the fundamental relation between work, heat and information~\cite{Maruyama2009,Toyabe2010,Eisert2015,Parrondo2015,Lutz2015,Quan2007,Uzdin2015}.
Here we apply the concept of the Szilard engine to a fully interacting quantum many-body system. 
We find that a working medium of a number of $N \geq 2$ bosons 
with attractive interactions is clearly superior to other previously discussed 
setups~\cite{Kim2011,Kim2012,Cai2012,Jeon2016,Zhuang2014,Lu2012}. 
In sharp contrast to the classical case, we find that the average work output increases 
with the particle number.  The highest overshoot occurs for a small but finite 
temperature, showing an intricate interplay between thermal and quantum effects.
We anticipate that our finding will shed new light on the role of information in controlling thermodynamic fluctuations in the deep quantum regime, which are strongly influenced by quantum correlations in interacting systems~\cite{Perarnau2015}. 
\end{abstract}

\maketitle

\section{Introduction}

The Szilard engine was originally designed as a thought experiment with only a single classical 
particle~\cite{Szilard1929} to illustrate the role of information in 
thermodynamics~(see, for example,~\cite{Parrondo2015} for a recent review).   
The apparent conflict with the second law could be resolved by 
properly accounting for the work cost associated with the information 
processing~\cite{Landauer1961,Bennett1982,Piechocinska2000,Plenio2001,Sagawa2008,Sagawa2009}.  
Although Szilard's suggestion dates back to 1929, only more recently the conversion between information and energy was shown experimentally using a Brownian particle~\cite{Toyabe2010}. 
A direct realisation of the classical Szilard cycle was reported by 
Rold{\' a}n {\it et al.}~\cite{Roldan2014}  for a colloidal particle in an optical double-well trap.  
 In a different scenario, Koski {\it et
  al.}~\cite{Koski2014,Koski2015} measured $k_B T \ln 2$ of work for
one bit of information using a single electron moving between 
two small metallic islands.
A quantum version of the single-particle Szilard engine was first
discussed by Zurek~\cite{Zurek1986}.  In contrast to the classical
case, insertion or removal of a wall in a quantum system shifts the
energy levels, implying that the process must be associated with
non-zero work~\cite{Bender2005, Gea2002, Gea2005}.  Kim {\it et
  al.}~\cite{Kim2011} showed that the amount of work that can be
extracted crucially depends on the underlying quantum statistics: two
non-interacting bosons were found superior to the classical
equivalent,  as well as to the corresponding fermionic case. \\
Many different facets of the quantum Szilard engine have been studied, including
optimisation of the cycle~\cite{Cai2012, Jeon2016} or the effect of
spin~\cite{Zhuang2014} and parity~\cite{Lu2012}, but all for
non-interacting particles.  The case of two attractive bosons was
discussed in Ref.~\cite{Kim2012}; however, the authors assigned the
increased work output to a classical effect. The question thus remains
how the information-to-work conversion in many-body quantum systems 
is affected by interactions between the particles. 

Here, we present a full quantum many-body treatment of spin-0 bosonic
particles in a Szilard engine with realistic attractive interactions between the particles, as they 
commonly occur in, for example, ultra-cold atomic gases~\cite{Bloch2008}. 
We  demonstrate quantum supremacy in the few-body limit for $N\le 5$, where a  solution to the full many-body problem can be obtained with very high numerical accuracy. 
A perturbative approach indicates that the supremacy further increases for larger particle numbers.
Surprisingly, the highest overshoot of work compared to $W_1=k_BT \ln 2$ ({\it i.e.}, the highest possible classical work output) occurs for a finite temperature, exemplifying the relation between thermodynamic fluctuations and the many-particle excitation spectrum.

\begin{figure*}[t]
\includegraphics[scale=0.55]{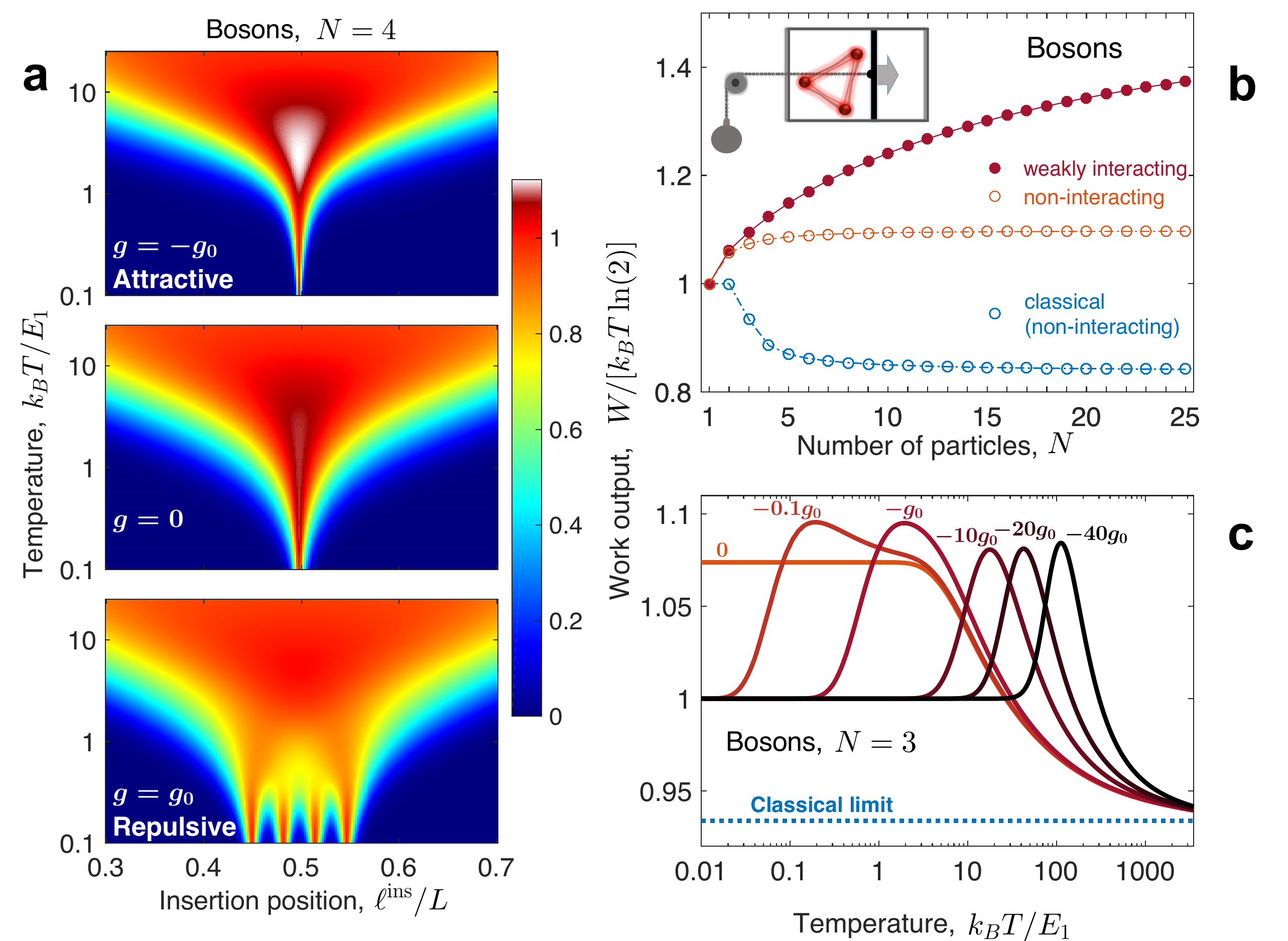}
\caption{{\bf Work output of the many-body Szilard
    engine.} {\bf a} For  $N=4$  bosons 
  the optimal work output  $W$ (in units of the
  classical single-particle work $W_1=k_B T \hbox{ ln 2}$) is found for
 attractive  interactions $g$ (upper panel)  
 at a finite temperature and around
 a symmetric insertion position of the barrier. 
(The interaction strength is in units of $g_0=\hbar^2/(Lm)$ with box length $L$ and 
single-particle mass $m$). The  optimal work output 
 exceeds the case of noninteracting (middle panel) and 
  repulsive bosons (lowest panel). Repulsive bosons exhibit $N$ peaks in the
  low-temperature limit for different insertion positions.  This
  behavior is similar to the case of non-interacting fermions and is a
  signature of the transition into a Tonks-Girardeau state (see text).
  {\bf b} The maximal work output $W/W_1$ increases significantly with
the  particle number for bosons with attractive interactions with $g=-0.01g_0$, 
   {\it (solid red line)}. It is always larger than the result for 
non-interacting  bosons {\it (dashed orange line)} and classical particles {\it
    (dashed blue line)}. 
  In each case, the temperature is chosen to
  maximise the relative work output. For non-interacting bosons, this
  occurs for $T\to 0$ (we used $k_BT/E_1=0.01$), while the interacting
  case is optimised at a finite temperature (like the white region in
  {\bf a}). For classical particles the result is independent of
  temperature.  Optimal insertion and removal positions of the wall
  are used to maximise $W/W_1$ for all considered systems. The inset
  shows a sketch of the many-particle Szilard engine performing work
  at the expansion step of the cycle.  {\bf c} The work output
  for $N=3$ as a function of temperature for different strengths of
  the attractive interaction. For large
  $T$, all curves converge into the classical result. }  
  \label{Fig:work_output}
\end{figure*}
%

\bigskip

\section{Many-body Szilard cycle for bosons with attractive interactions.} 
  Our claim is based on a fully {\it ab
  initio} simulation of the quantum many-particle Szilard cycle by
exact numerical diagonalisation, {\it i.e.}, the full configuration
interaction method (as further described in the supplementary material).  
A hard-walled one-dimensional container of length $L$ confines $N$ bosons that constitute
the working medium. We model the interactions by the usual two-body pseudopotential of 
contact type~\cite{Bloch2008}, $g\delta (x_1-x_2)$, where the strength of the interaction $g$ is
given in units of $g_0=\hbar^2/(Lm)$.  The single-particle ground
state energy $E_1=\hbar^2\pi^2/2mL^2$ sets the
energy unit, where $m$ is the mass of a single particle. 
The cycle of the Szilard engine goes
through four steps, assumed to be carried out quasi-statically and in
thermodynamic equilibrium with a single surrounding heat bath at 
temperature $T$: (i) insertion of a wall dividing the quantum
many-body system at a position $\ell^{\mathrm {ins}}$, followed by
(ii) a measurement of the actual particle number $n$ on the left side
of the wall, (iii) reversible translation of the wall to its final
position $\ell_n^{\mathrm{rem}}$ depending on the outcome $n$ of the
measurement, and finally (iv) removal of the barrier at 
$\ell_n^{\mathrm{rem}}$.

The total average work output of a single cycle with processes
(i)-(iv) has  been determined~\cite{Kim2011} as 
\begin{equation}
W = -k_BT \sum^N_{n = 0} p_n(\ell^{\mathrm{ins}}) \ln \left[ \frac{p_n(\ell^{\mathrm{ins}})}{p_n(\ell^{\mathrm{rem}}_n)} \right] ~.\label{Eq:Wtot}
\end{equation}
Here, $p_n(\ell)$ denotes the probability to find $n$ particles to
 the left of the wall located at position $\ell$, and $N-n$
particles to the right, if the combined system is in thermal
equilibrium. The $N$-particle
eigenstates $\Psi_i$ with energy $E_i$, 
obtained by numerical diagonalisation, can be classified by the particle
number $n_i$ in the left subsystem with $0<x<\ell$. 
Then we find  that $p_n(\ell)=\sum_i\delta_{n_i,n} e^{-E_i(\ell)/k_BT}/Z$ with $Z=\sum_i
e^{-E_i(\ell)/k_BT}$.

Measuring the particle number on one side after insertion
of the wall, one gains the Shannon
information~\cite{ShannonBellSystemTechJ1948}
$ I=-\sum_{n=0}^N p_n(\ell^\mathrm{ins}) \ln p_n(\ell^\mathrm{ins}).
  \label{EqImeasuerment} $
Going back to the original state in the cycle, this information is
lost, associated with an average increase of entropy $\Delta
S=k_BI$. This increase in entropy of the system allows one to extract the
average amount of work $W\le k_BT I$ which can be positive.
Here, the equality only holds if all $p_n(\ell^{\mathrm{rem}}_n)\equiv
1$. In this case the removal of the barrier is reversible for each
observed particle number. This reversibility had been associated with
the conversion of the full information gain into
work~\cite{Horowitz2011}, as explicitly assumed in
Ref.~\cite{Plesch2014}. While $p_n(\ell^{\mathrm{rem}}_n)\equiv 1$ is
straightforward for the single-particle case with
$\ell^{\mathrm{rem}}_0=0$ and $\ell^{\mathrm{rem}}_1=L$, this is hard
to realise for $N\ge 2$~\cite{Horowitz2011}.  

For our case of a moving piston, the full work can typically not be extracted.  To
optimise $W$, we choose the optimal $\ell^{\mathrm{rem}}_n$,
maximising $p_n(\ell^{\mathrm{rem}}_n)$ for all systems considered here. 
(The procedure is similar to the non-interacting case~\cite{Jeon2016}).

The highest relative work output is obtained for a many-body system of
{\it attractive} bosons at a finite temperature.
This is the white region in the top panel of
Fig.~\ref{Fig:work_output} (a), where the work output $W/W_1\approx 1.12$ 
for a system of four attractive bosons surpasses the
results for noninteracting (middle panel with $W/W_1\lesssim 1.08$) 
as well as for repulsive (lowest panel) bosons. For comparison a system of
four classical particles has $W/W_1\lesssim 0.886$ (not shown here).
We also note that for interaction strengths $g\le 0$, the maximum work
output always occurs if the wall is inserted in the middle of the
container ($\ell^\textrm{ins}=L/2$) for an engine operating in the
deep quantum regime. (For larger temperatures, other insertion positions can become favorable, see the Supplementary Material). For $T\to 0$, the work output vanishes 
 if $\ell^\textrm{ins}\neq L/2$. In this limit all non-interacting
bosons occupy the lowest single-particle quantum level. After
insertion of the wall, the energetically lowest-lying level is in the
larger region. For $\ell^{\text{ins}} \neq L/2$ we know beforehand the
location of the particles and measuring the number of particles does
not provide any new information, {\it i.e.}, $I=0$. Consequently, no
work can be extracted in the cycle. Attractive interactions obviously
enhance this feature.  However, this does not hold for repulsive
interactions, $g>0$, as shown in the lowest panel of
Fig.~\ref{Fig:work_output} (a). In this case, the particles spread out
on different sides of the wall in the ground state.  Here,
degeneracies between different many-particle states occur at
particular values of $\ell^\textrm{ins}$, which allow an information
gain in the measurement. This explains the $N$ distinct peaks as a
function of $\ell^\textrm{ins}$ for low temperature in the lowest panel of Fig.~\ref{Fig:work_output} (a).

The maximum of $W/W_1$ for attractive bosons  increases with
particle number, as shown in Fig.~\ref{Fig:work_output} (b). The
optimal relative work output  is higher for
attractive bosons ({\it solid red line}) than for non-interacting
bosons ({\it red dashed line}) and clearly beats the corresponding
result for classical particles ({\it blue dashed line}). 
Here, the data for $N\le 5$ were obtained by exact diagonalisation  while a 
perturbative approach (see supplementary material) was applied for $N>5$.
The peak work output
for bosons with attractive interactions at a finite temperature is a
general feature, which holds for a wide range of interaction strengths,
see Fig.~\ref{Fig:work_output} (c) for the case of $N=3$ bosons.  Indeed,
the temperature at which the peak occurs increases with larger interaction
strengths. 

\bigskip

\section{Onset of the peak at an intermediate temperature.} 

For systems with attractive interactions,  $g<0$, 
the work output equals $k_BT \ln 2$ at low temperatures,  independent of $N$. 
Due to the dominance of the attractive interaction, all $N$ particles will be
found on one side of the barrier. When the barrier is inserted symmetrically, 
we have $p_0(L/2)=p_N(L/2)=1/2$, while all other
$p_n(L/2)=0$. At the same time, the removal position
$\ell_0^\textrm{rem}=0$ and $\ell_N^\textrm{rem}=L$ provide
$p_0(\ell_0^\textrm{rem})=1$ and $p_N(\ell_N^\textrm{rem})=1$, so that
Eq.~(\ref{Eq:Wtot}) provides the work output $W=k_BT \ln 2$ for the
entire cycle as observed in Fig.~\ref{Fig:work_output}(c). This case, with two
possible measurement outcomes and a full sweep of the piston,
resembles the single-particle case. One might wonder, whether the
increased particle number should not imply a higher pressure on the
piston and thus, more work. This, however, is not the case, as the
attraction between the particles reduces the pressure. 
Also, when inserting the barrier, the
difference in work due to the interactions has to be taken into account.
With increasing temperature ({\it i.e.}, $k_BT\sim -3g(N-1)/L\approx
-0.6(N-1)E_1g/g_0$, for weak interactions as shown in the supplementary material) other
measurement outcomes than $n=0$ or $n=N$ become probable. Since $p_0$ and
$p_N$ now decrease with temperature we see a deviation from the
performance of the single-particle engine.

\section{The two-particle interacting engine.}

To get a better understanding of the physics behind the enhancement of work output for
bosons with attractive interactions at finite temperatures, let us look at the two-particle
case in some more detail. For a central insertion of the barrier, we find
$p_0(L/2) = p_2(L/2)$.   For the same symmetry reasons,
$p_1(\ell )$ has a maximum at this barrier position. No work can thus be
extracted in cycles where the two particles are measured on different
sides of the central barrier, since 
$p_1(\ell ^{\mathrm {ins}}) / p_1(\ell _1^{\mathrm {rem}})\ge 1 $ in Eq.~(\ref{Eq:Wtot}). 
\begin{figure}[t]
\includegraphics[scale=0.6]{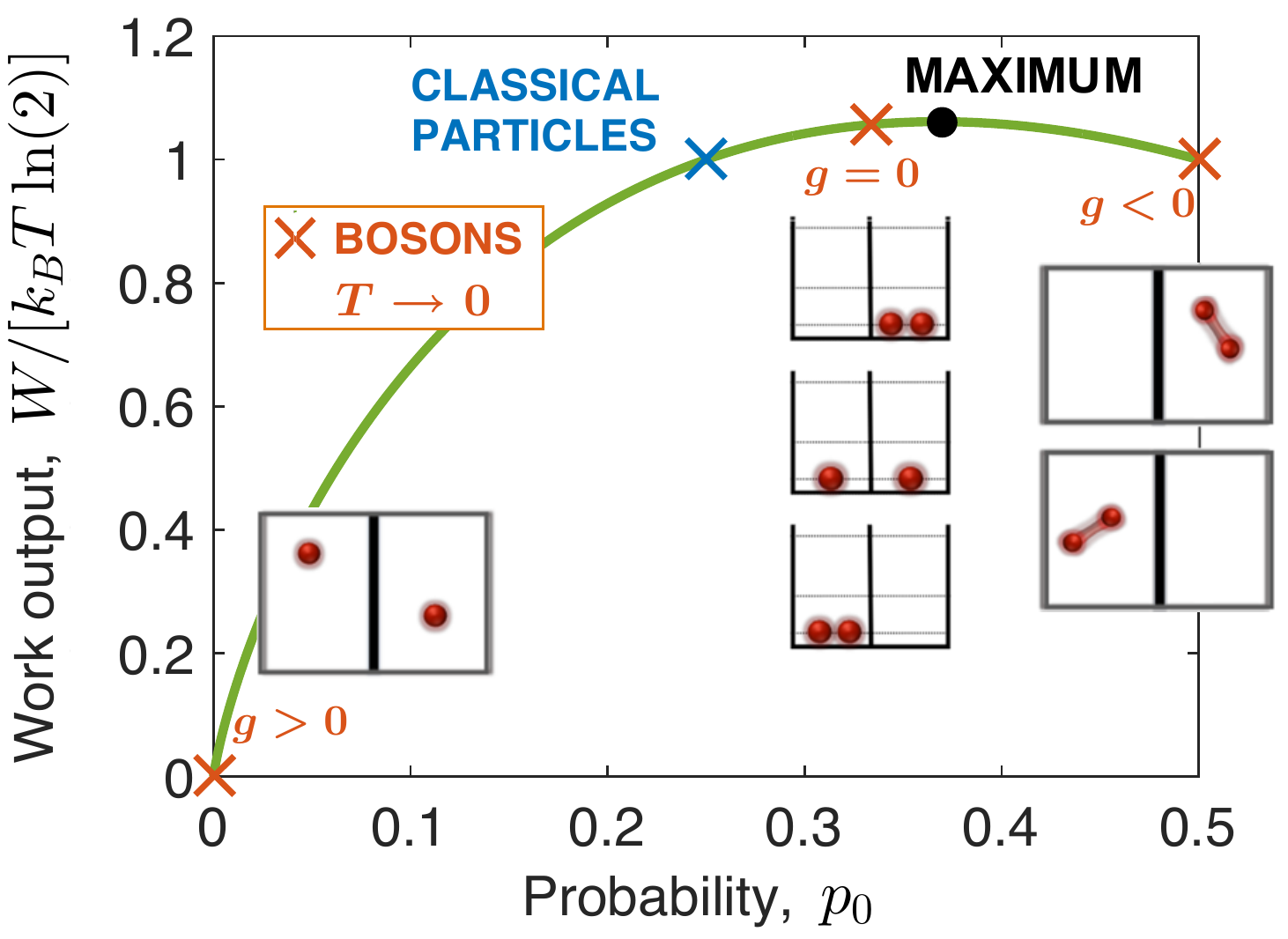}
\caption{{\bf Work output per cycle for the two-particle Szilard engine.}
For a symmetric insertion of the barrier the work output depends solely  on the probability $p_0$ to find all particles on the right side. For classical particles, $p_0=1/4$ holds, and this result is also reached in all other cases in the limit of high temperatures. However, in the limit of low temperatures,  $p_0$ differs for bosons with different interactions. The insets show the two-particle configurations in the repulsive case (left) and the attractive case (right), as well as the level degeneracy for non-interacting bosons.}
\label{Fig:Two_Particle}
\end{figure}
Thus, the only contributions to the work output  result  from
 $p_0$ and $p_2$. Together with $p_0(\ell^\textrm{rem}_0=0) = p_2(\ell^\textrm{rem}_2=L) = 1$ we obtain
\begin{equation}
W = - 2k_B T p_0(L/2)\ln p_0(L/2) \label{Eq:W2}
\end{equation}
This function has its peak at $p_0 = 1/e$ with the peak value
$W \approx 1.061 k_BT \ln 2$, see Fig.~\ref{Fig:Two_Particle}. 
This implies a finite value $p_1=1-2/e$. Even if no work can be
extracted with one particle on either side of the barrier, a non-zero
probability $p_1$ of such a measurement outcome can be preferable. 

Two attractive bosons, initially at $T\rightarrow 0$ and with $p_0=1/2$,
will for increasing $T$ continuously approach the classical limit of
$p_0=1/4$.  Hence, at a certain temperature, depending on the
interaction strength, $p_0$ passes through $p_0=1/e$ producing a peak in
the relative work. Physically, one may understand this property of the
engine as follows: At low temperatures, the two attractive bosons will
always end up on the same side of the barrier,  bound together by their attraction.  
The cycle is then operating similar to the single
particle case, which explains that $W = k_BT \ln 2$ when 
$p_0 = 1/2$. A less correlated system (obtained with increasing $T$) provides
a larger expansion work for cycles in which both particles are on one
side of the barrier. On the other hand, cycles with one particle on
each side of the barrier, from which no work can be extracted, become
more frequent.   
For $1/e < p_0 < 1/2$, the enhanced pressure is more important and the average work output
increases with decreasing $p_0$. 
For lower values of $p_0$, i.e. $p_0 < 1/e$, too few cycles 
contribute on average to the work production. The average work output decreases with decreasing $p_0$ despite the corresponding increase in pressure.  
Importantly, we note the absence of a similar maximum in the non-interacting case, 
where $W/W_1$  is found to decrease steadily towards the classical limit with increasing $T$.

\section{Szilard engines with $N>2$ attractive bosons.}

\begin{figure}[ht]
\includegraphics[scale=0.45]{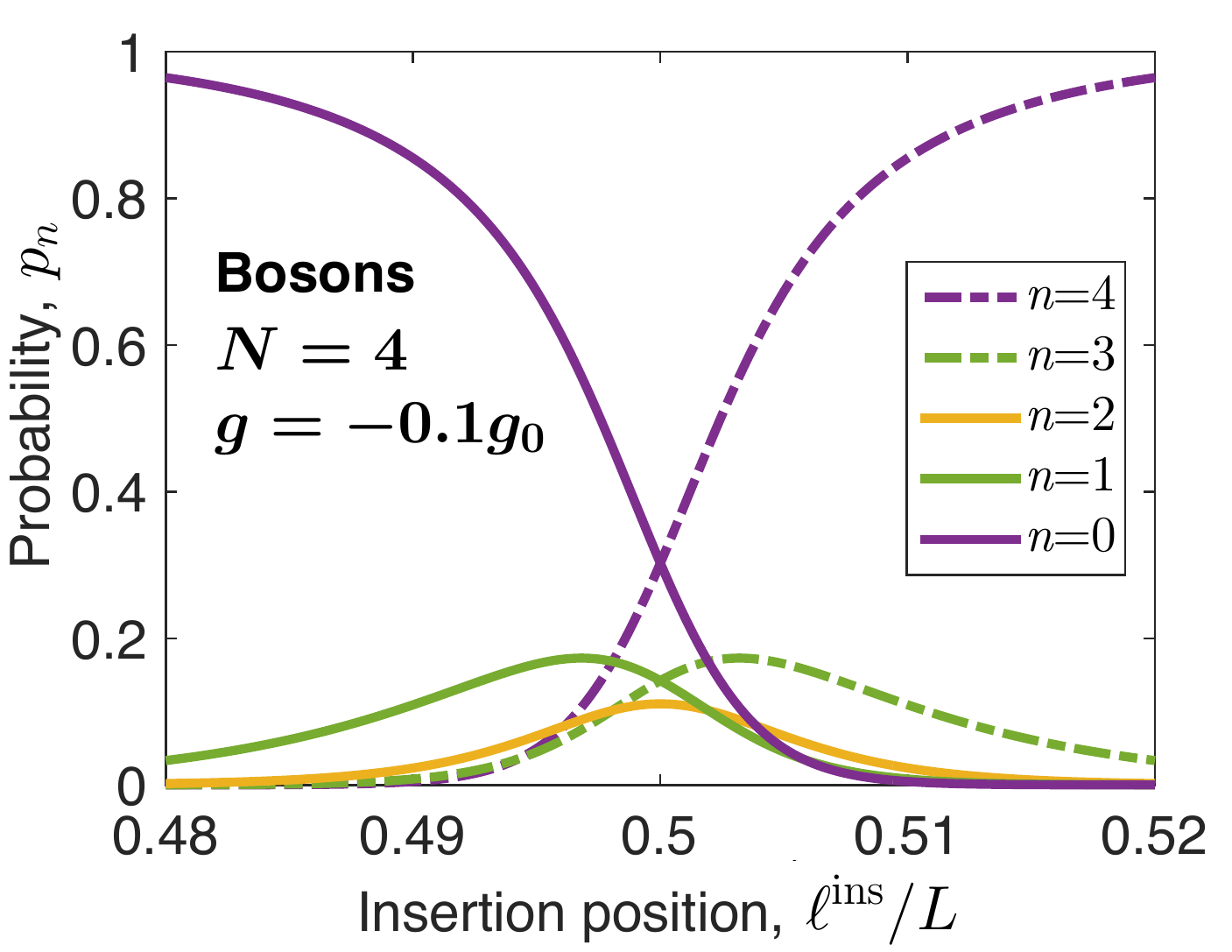}
\caption{{\bf Probability distributions.} 
Work can be extracted in all cycles with insertion of the wall at the midpoint, except for the case with equally many particles on either side. Shown are the 
probability distributions assigned to the different measurement outcomes at the temperature of the maximal relative work output ($k_BT/E_1\approx 0.243$) for  $N=4$ bosons with weak attraction, $g=-0.1g_0$. 
} \label{Fig:Insertion_Position}
\end{figure}

The maximum of $W/W_1$ tends to increase with the particle number $N$ (as previously discussed in connection with Fig.~\ref{Fig:work_output}~b). The reason lies in the fact that work can be extracted from a larger number of  measurement outcomes.    
Similar to the two-particle engine, the combined contribution to the average work 
output from cycles in which all
particles are on the same side of a barrier inserted at
$\ell^{\text{ins}}=L/2$ is given by Eq.~(\ref{Eq:W2}). However, also cycles with $n=1,2,\ldots,N-1$
(except if $n = N/2$) on the left side of the barrier do contribute to
the average work output, and work output  even higher than in
the two-particle case is possible. 
The maximum of $p_1(\ell)$ and that of $p_{N-1}(\ell)$ occurs for $\ell \neq L/2$, as clearly indicated
by the probabilities for different measurement outcomes shown for $N=4$ in 
Fig.~\ref{Fig:Insertion_Position}.  This means that 
$p_n(\ell ^{\mathrm{ins}}) /p_n (\ell _n^{\mathrm{rem}})\le 1$ 
is possible for $\ell ^{\mathrm{ins}}=L/2$ and that  work may be extracted in agreement with Eq.~(\ref{Eq:Wtot}).
For all systems considered here, 
with  insertion of the barrier at the midpoint the optimum is reached for $p_0 = p_N \approx 0.3$ (see the example for $N=4$ in Fig.~\ref{Fig:Insertion_Position}), which is close to the
optimal value of $1/e$ for the corresponding two-particle engine. 

\section{Repulsive Bosons}

Finally, we consider the repulsive interactions between bosons, see
Figure \ref{Fig:work_output}(c). In the low-temperature
limit, the   relative work output is very similar to that of
non-interacting spin-less fermions discussed in  Refs.~\cite{Kim2012,
  Cai2012}. This resemblance becomes even more pronounced with
increasing interaction strength. This in fact is no coincidence, but
rather a property of one-dimensional bosons with strong, repulsive
interactions that have an impenetrable core: Indeed, in the limit of
infinite repulsion, bosons act like spin-polarised non-interacting
fermions. This is the well-known Tonks-Giradeau regime~\cite{Girardeau1960}.  
Both for non-interacting fermions and strongly repulsive bosons, the region
where the quantum Szilard engine exceeds the classical single-particle maximum of work
output, has disappeared.

\section{Conclusions}

We have demonstrated that the work output of the quantum Szilard engine can be significantly boosted by short-ranged attractive interactions for a bosonic working medium. We based our claim on the (numerically) exact solution of the full many-body  Schr\"odinger equation for up to five bosons. 
It is likely that the effect is even further enhanced for larger particle numbers; however, despite the simple one-dimensional setup, the numerical effort grows very significantly (and beyond our feasibility) for larger $N$.
By increasing the strength of the interparticle attraction, the engine's work output can be increased significantly also at higher temperatures, where the work that can be extracted generally is of larger magnitude. While we here restrict our analysis to idealised quasi-static processes, it would be of much interest 
to consider a finite speed in the ramping of the barrier, enabling transitions to excited states which by coupling to baths will lead to dissipation. Extending our approach to quantify irreversibility in real processes on the basis of a fully {\it ab initio} quantum description may in the future allow to study
dissipative aspects in the kinetics of the conversion between information and work.

\bibliographystyle{naturemag_noURL}
\bibliography{BENGTSSON_SZILARD}

\newpage
\newpage

{\centerline 
{\bf Supplementary Material}
}

\bigskip

\noindent
{\bf 1. Work output of the quantum Szilard engine}

\medskip

In contrast to other conventional heat engines that operate by
exploiting a temperature gradient, as discussed in many textbooks on
thermodynamics, the Szilard engine~\cite{Szilard1929} 
allows for work to be extracted also when
connected to a {\it single} heat bath at constant temperature.
It is propelled by the information obtained about the working
medium and its microscopical properties.
In the supplementary material, we briefly outline the theoretical description of the quantum
Szilard engine, in close analogy to that of Refs.~\cite{Kim2011, Jeon2016}. 
An idealised version of the Szilard engine cycle consists of four
well-defined steps: {(i) \it 
  insertion}, {(ii) \it measurement}, {(iii) \it expansion} and finally
{(iv) \it removal}. First, an impenetrable barrier is introduced 
  (i) that effectively splits the working medium into two
halves. Then, the number of particles on each side of the barrier is
measured (ii). Depending on the outcome of this measurement, the
barrier moves (iii) to a new position and contraction-expansion
work can be extracted in the process. Finally, the barrier is removed
(iv) which completes a single cycle of the engine.

All four steps, (i)-(iv), of the Szilard engine are assumed to
be carried out quasi-statically and in thermodynamic equilibrium with
a surrounding heat bath at temperature $T$. Now, the work associated
with an isothermal process can be obtained from
\begin{equation}
 W \leq -\Delta F = k_B T \Delta \left( \ln Z\right), \label{Wdef}
\end{equation}
where $k_B$, $F$ and $Z$ are the Boltzmann constant, the Helmholtz free
energy and the partition function
\begin{equation}
Z = \sum_j e^{-E_j/(k_BT)}, \label{Z}
\end{equation}
where the sum runs over the energies $E_j$ of, in principle, all micro
(or quantum) states of the considered system. In practice, however, we
construct an approximate partition function from a finite number of
energy states. Note that the work in Eq.~(\ref{Wdef}) is chosen
to be positive if done {\it by} the system. Also, the equivalence between
$W$ and $-\Delta F$ is reserved for reversible processes alone.

We now turn to the work associated with the individual steps of the
quantum Szilard engine. For simplicity, we consider an engine with $N$
particles initially confined in a one-dimensional box of size $L$. All
steps of the engine are, as previously mentioned, carried out
quasi-statically and in thermal equilibrium with the surrounding heat
bath at temperature $T$. To maximise the work output, we further
assume that all involved processes are reversible, unless specified
otherwise.

{(i) \it  Insertion}. A  wall is slowly introduced at $\ell^{\mathrm{ins}}$, 
where $0 \leq \ell^{\mathrm{ins}} \leq L$. In the end, the initial system is divided
into left and right sub-systems of sizes $\ell^{\mathrm{ins}}$ and $L-\ell^{\mathrm{ins}}$
respectively. Based on Eq.~(\ref{Wdef}), the work of this process
is given by
\begin{equation}
W_{\rm (i)} = k_BT \ln \biggl [ \frac{\sum_{n=0}^{N} Z_{n}(\ell^{\mathrm{ins}})}{Z_N(L)}  \biggr ], 
\label{Win}
\end{equation}
where $Z_n(\ell^{\mathrm{ins}})$ is the short-hand notation for the
partition function obtained with $n$ particles in the left sub-system
and $N-n$ in the right one. With this notation, $Z_N(L)$ is thus
equivalent to the partition function of the initial system, before the
insertion of the barrier. Also, note that prior to measurement, the
number of particles on either side of the barrier is not yet a
characteristic property of the new system. We need thus to sum over
all possible particle numbers in the numerator of
Eq.~(\ref{Win}). Finally, we want to stress the fact that, unlike for
a classical description of the engine, the insertion of a barrier
costs energy in the form of work, due to the associated change in the
potential landscape.

{(ii) \it Measurement}. The number of particles located on the different
sides of the barrier is now measured. Here, following ~\cite{Kim2012}, we assume that the
measurement process itself costs no work, {\it i.e.}, we assume that
$W_{\rm (ii)}=0$ (see main text). The probability that $n$ particles are measured to be
on the left side of the barrier (and $N-n$ on the right side) is given by
\begin{equation}
p_n(\ell^{\mathrm{ins}}) = \frac{Z_n(\ell^{\mathrm{ins}})}{\sum_{n^{\prime}=0}^N Z_{n^{\prime}}(\ell^{\mathrm{ins}})}. \label{pn}
\end{equation}

{(iii) \it Expansion}. The barrier introduced in {(i)} is assumed to move without
friction. During this expansion/contraction process, the number of
particles on either side of the barrier remains fixed. In other words,
the barrier is assumed high enough such that tunnelling may be
neglected. If the barrier moves from $\ell^{\mathrm{ins}}$ to
$\ell^{\mathrm{rem}}_n$ when $n$ particles are measured in the left
sub-system, the average work extracted from this step of the cycle
reads
\begin{equation}
W_{\rm (iii)}= k_BT \sum_{n=0}^N p_n(\ell^{\mathrm{ins}}) \ln \left[\frac{Z_{n}(\ell^{\mathrm{rem}}_n)}{Z_{n}(\ell^{\mathrm{ins}})}  \right], \label{Wexp}
\end{equation}
where $p_n$ are the probabilities given by Eq.~(\ref{pn}).

{(iv) \it Removal}. The barrier at $\ell^{\mathrm{rem}}_n$, that
separates the left sub-system with $n$ particles from the right one
with $N-n$ particles, is now slowly removed. As the height of the barrier shrinks, particles will
eventually start to tunnel between the two sub-systems. This transfer
of particles makes the removal of the barrier an irreversible
process. Clearly, if we instead were to start without a barrier and
introduce one at $\ell^{\mathrm{rem}}_n$, then we can generally not be
certain to end up with $n$ particles to the left of the
partitioner. Assuming that the particles are fully delocalised between
the two sub-systems already in the infinite height barrier limit, then
the average work associated with the removal process is given by
\begin{equation}
W_{\mathrm{(iv)}} = k_B T \sum_{n=0}^N p_n(\ell^{\mathrm{ins}}) \ln
\left[\frac{Z_{N}(L)}{\sum_{n^{\prime}=0}^N
    Z_{n^{\prime}}(\ell^{\mathrm{rem}}_n)} \right]. \label{Wrem}
\end{equation}

Finally, the averaged combined work output of a single Szilard cycle, $W$, is given by the sum
of the partial works associated with the four steps {(i)-(iv)}, i.e. 
$W=W_{\rm (i)}+W_{\rm (ii)}+W_{\rm (iii)}+W_{\rm (iv)}$, and simplifies into
\begin{equation}
W = -k_BT \sum^N_{n = 0} p_n(\ell^{\mathrm{ins}}) \ln \left[ \frac{p_n(\ell^{\mathrm{ins}})}{p_n(\ell^{\mathrm{rem}}_n)} \right]. \label{Wtot}
\end{equation}
which is the central equation (1) in the main article.

\bigskip

\noindent
{\bf 2. The interacting many-body Hamiltonian and exact diagonalisation} 
 
\medskip
 
 To keep the schematic setup of the many-body Szilard cycle as simple as possible, we consider a quantum system of $N$ interacting particles, initially confined in a one-dimensional box of size $L$ that is separated by a barrier inserted at a certain position $\ell $. 
We note that for contact interactions between the particles, as defined in the main text, the exact energies $E_j$ to the fully interacting many-body Hamiltonian $\hat H$ are those given in terms of two independent  systems with $n$ and $N-n$ particles.
In order to construct the partition functions and compute the probabilities $p_n$, the entire exact many-body energy spectrum is needed. For the simple case of non-interacting particles (or single-particle systems) these energies are known analytically. For interacting particles, however, they must be determined by solving the full many-body problem. We here apply 
the configuration interaction method where we use a basis of the 5th order B-splines~\cite{deBoor2001}, with a linear distribution of knot-points within each left/right sub-system, to determine the energies of each sub-system and parity at each stage.  For $N = 3$, we used 62 B-splines (or one-body states) to construct the many-body basis for each sub-system. Since the dimension of the many-body problem grows drastically with $N$, we needed to decrease the number of B-splines to 32 for $N=4$. Consequently, in this case we could not go to equally high temperatures and interaction strengths. 

\bigskip

\noindent
{\bf 3. Perturbative approach for weakly attractive bosons at 
low temperatures}

\medskip

Here we consider the case $k_BT\ll E_1$, so that only the lowest quantum
levels in each part are thermally occupied. In the case of vanishing
interaction, the state with $n$ particles in the lowest level of the left
part of the wall positioned
at $\ell$ (and $N-n$ particles in the lowest level of the right
part) has the energy
\[
E_n^{(0)}(\ell)=n E_1\left(\frac{L}{\ell}\right)^2+
(N-n)E_1\left(\frac{L}{L-\ell}\right)^2
\]
Applying the wave function $\Psi_0(x)=\sin (\pi
x/\ell)\sqrt{2/\ell}$ for the left side, 
the mutual interaction energy between two particles in this
level is 
\begin{equation}
U(\ell)=g\int_0^{\ell} dx\, |\Psi_0(x)|^4=\frac{3g}{2\ell}
\end{equation}
Now we assume that 
this interaction  energy (times the number of interacting partners) 
is much smaller than the level spacing, i.e. $(n-1)U(\ell)\ll
3 E_1\left(\frac{L}{\ell}\right)^2$, which is satisfied for
$g\ll \frac{\pi^2 g_0}{N-1}$. Then we may determine
the energy of the many-particle state  by first-order perturbation theory. 
This results in the interaction energy
\begin{equation}
E_n^{(1)}(\ell)=\frac{n(n-1)}{2}U(\ell)+\frac{(N-n)(N-n-1)}{2}U(L-\ell)
\, .\label{Eq:Eapprox}
\end{equation}
Setting $E_n(\ell)\approx E_n^{(0)}(\ell)+E_n^{(1)}(\ell)$, we obtain an analytical
expression for the probabilities $p_n(\ell)$ without any need for numerical
diagonalisation. 
Using $\ell^\mathrm{ins}=L/2$ and
determining the optimal removal positions $\ell^\mathrm{rem}_n$ numerically,
we get the work output by Eq.~(\ref{Eq:Wtot}). Again the optimal
temperature needs to be chosen to obtain the results plotted in
Fig.~\ref{Fig:work_output}(b). \\

\noindent
{\bf 4. Estimate of the peak temperature} 

For the symmetric wall position, the
ground state of the system with attractive bosons has all particles on one
side, say the left one. Using the perturbative approach discussed above, the
interaction energy is  $E_N^{(1)}(L/2)$. If one boson is transferred from the
left side to the right side, the interaction energy changes to
$E_{N-1}^{(1)}(L/2)$, while the level energies $E_n^{(0)}(L/2)$ are
independent on $n$ for the  symmetric wall position. Thus thermal excitations
become likely for $k_BT\sim E_{N-1}^{(1)}(L/2)-E_N^{(1)}(L/2)=-3(N-1)g/L$. For
these temperatures the  particles do not cluster on the same side of the wall
any longer and we have $p_0<1/2$.

\noindent
{\bf 5. Temperature dependence of the work output for different interaction
  strengths}
  
\begin{figure}[ht]
\centering
\includegraphics[scale=0.5]{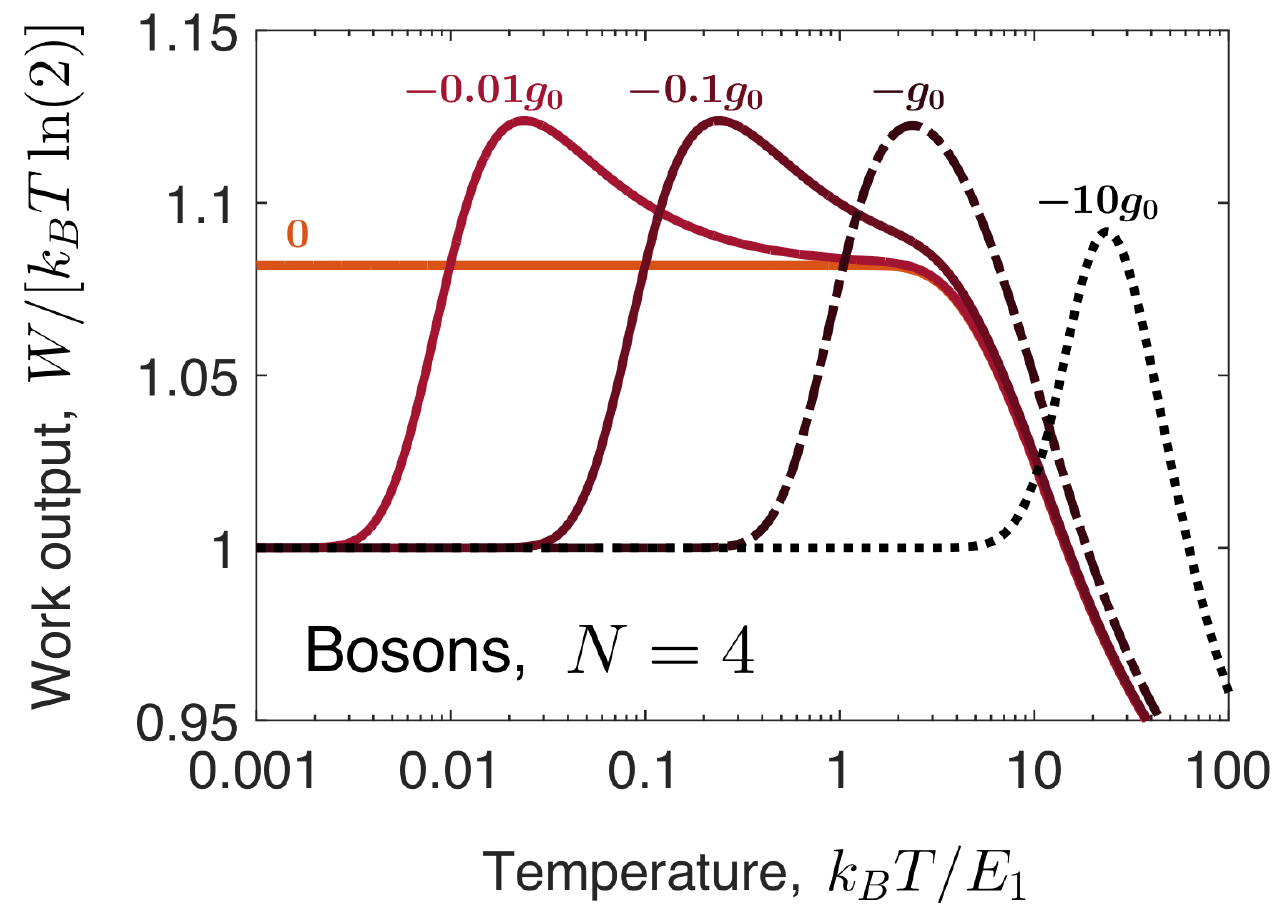}
\caption{{\bf Temperature dependence of the work output for $N=4$ bosons.} 
With increasing interaction strength, the peak in 
the relative work output occurs at a higher temperature, similar to what was shown in Fig.~1(c) for $N=3$. 
}
\label{Fig:N4temp}
\end{figure}

As a complement to Fig.~1(c) of the main article, we show the case for $N=4$
here in Fig.~\ref{Fig:N4temp}.
For small to medium couplings $-g_0\lesssim g <0$, 
the peaks have approximately the same
height and they are shifted proportionally to $g$. This shift follows the
deviations from the low temperature limit $W=W_1$, which set in 
at $k_BT\approx -0.6(N-1)E_1g/g_0$, as shown by the approximative approach in 
the main article. 
As discussed in the method section, for $g\approx -\pi^2 g_0/(N-1)$,
correlation effects become important and we find a reduced peak at $g=-10
g_0$, similar to the case for $N=3$ in Fig.~1(c) of the main article. 
Due to the high numerical demand on the 
numerical diagonalization, we did not obtain results for larger $|g|$ in 
the case $N=4$, while for $N=3$ an increase of the peak height for even larger
$|g|$ is observed. 

For all interaction strengths $g<0$, the peak height
is actually larger than the peak for the attractive two-particle case
$W_2\approx 1.061k_BT\ln(2) $ depicted in Fig.~2 of the main article. This is
due to the fact, that Eq.~(2) of the main article is a lower bound for the
work output and $p_0(L/2)$ necessarily moves from $1/2$ at $T\to 0$ to the
classical result $1/2^N$ at large temperatures. Thus the maximum for $p_0=1/e$
is taken at some intermediate temperature.

\noindent
{\bf 6. Operation of the Quantum Szilard engine at high temperatures}

\begin{figure}
\includegraphics[scale=0.45]{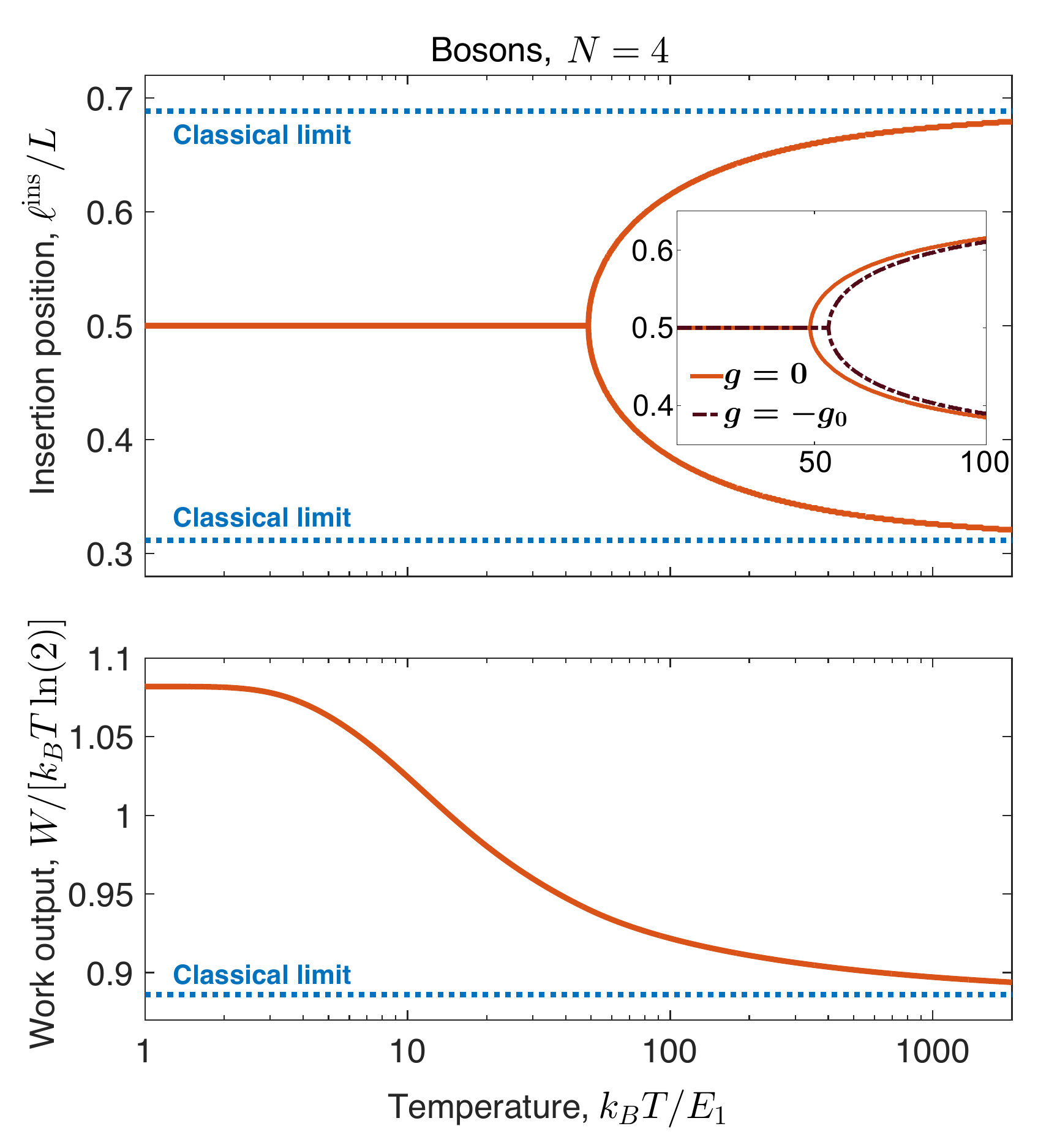}
\caption{
        {\bf Pitchfork bifurcation for the optimal insertion position at a
          critical temperature for  $N=4$ bosons}.  For large temperatures, the optimal insertion
        position becomes asymmetric in order to recover the classical result (blue
        dashed line). The full red line shows the case without interactions
        $g=0$. The inset in the upper panel shows the corresponding result for 
        $g=-g_0$ (dark-red lines).}
\label{Fig:Bifurcation}
\end{figure}

For $N=4$ particles, Fig.~\ref{Fig:Bifurcation} shows that the symmetric
insertion point $\ell^\mathrm{ins}=L/2$ is not optimal for high temperatures. 
For classical particles, the optimal work output is given by 
\begin{align}\label{cszewtot}\nonumber
W_{\mathrm{tot}} &= -N\kb T\left[\left(\frac{\xins}{\lt}
\right)^N\ln\left(\frac{\xins}{\lt} \right)  \right. \\\nonumber
&+\left.\left(1-\frac{\xins}{\lt}
\right)^N\ln\left(1-\frac{\xins}{\lt}\right)\right]\\\nonumber
& - \kb T\sum_{m=1}^{N-1}\binom{N}{m}
\left(\frac{\xins}{\lt}\right)^m \left(1-\frac{\xins}{\lt} \right)^{N-m} \\
&\times \ln\left[\frac{\left(\frac{\xins}{\lt}\right)^m
\left(1-\frac{\xins}{\lt}\right)^{N-m}}
{ \left(\frac{m}{N}\right)^m \left(1-\frac{m}{N}\right)^{N-m}} \right].
\end{align}
A numerical scan of different insertion positions shows that an
asymmetric insertion point $\ell^\mathrm{ins}\ne L/2$ (as shown by the blue
dashed line in Fig.~\ref{Fig:Bifurcation}) provides the highest work output.
In contrast, the symmetric position is optimal for  $N\le 3$
classical particles. For the non-repulsively interacting bosons  ($g\le 0$)
studied here, the symmetric insertion is favorable in the low temperature limit as thoroughly discussed in
the main article. On the other hand, for large temperatures the classical
result needs to be recovered. This occurs via a pitchfork bifurcation\cite{GuckenheimerBook1983} at an
intermediate temperature as shown in Fig.~\ref{Fig:Bifurcation}.
For noninteracting Bosons it occurs at $k_BT_c\approx 50 E_1$ for $N=4$, and
at slightly larger values if attractive interactions are included. 

\vfill\eject
\end{document}